\documentclass[letterpaper, 10 pt, conference]{ieeeconf} 

\IEEEoverridecommandlockouts
\let\labelindent\relax
\usepackage{cite}
\usepackage{amsmath,amssymb,amsfonts,amsthm}
\usepackage{algorithmic}
\usepackage{graphicx} 
 \usepackage{floatrow}
\usepackage{textcomp}
\usepackage[dvipsnames]{xcolor}
\usepackage{dsfont}
\usepackage[hidelinks]{hyperref}
\usepackage{multirow}
\usepackage{bbm}
\usepackage[version=4]{mhchem}
\usepackage{siunitx}
\DeclareMathOperator*{\sat}{sat} 
\usepackage{verbatim}
\usepackage{bm}
\usepackage{framed}
\usepackage[ruled,linesnumbered]{algorithm2e}
\usepackage{savefnmark}
\usepackage{pifont}
\usepackage{enumitem}
\newcommand{\va}{{\bf a}}
\newcommand{\vb}{{\bf b}}
\newcommand{\vd}{{\bf d}}
\newcommand{\ve}{{\bf e}}
\newcommand{\vf}{{\bf f}}
\newcommand{\vg}{{\bf g}}
\newcommand{\vl}{{\bf l}}
\newcommand{\vq}{{\bf q}}
\newcommand{\vs}{{\bf s}}
\newcommand{\vu}{{\bf u}}
\newcommand{\vv}{{\bf v}}
\newcommand{\vx}{{\bf x}}

\newcommand{\vA}{{\bf A}}
\newcommand{\vB}{{\bf B}}
\newcommand{\vC}{{\bf C}}
\newcommand{\vE}{{\bf E}}
\newcommand{\vF}{{\bf F}}
\newcommand{\vG}{{\bf G}}
\newcommand{\vH}{{\bf H}}
\newcommand{\vJ}{{\bf J}}
\newcommand{\vK}{{\bf K}}
\newcommand{\vM}{{\bf M}}
\newcommand{\vU}{{\bf U}}
\newcommand{\vV}{{\bf V}}
\newcommand{\vR}{{\bf R}}
\newcommand{\vX}{{\bf X}}

\newcommand{\vsigma}{\ensuremath{\bm{\sigma}}}
\newcommand{\vxi}{\ensuremath{\bm{\xi}}}
\newcommand{\vomega}{\ensuremath{\bm{\omega}}}
\newcommand{\vzeta}{\ensuremath{\bm{\zeta}}}

\newcommand{\TdeltaT}{\ensuremath{T_{\delta_T}}}
\newcommand{\deltaT}{\ensuremath{\delta_T}}
\newcommand{\deltaTBL}{\ensuremath{\delta_{T_{BL}}}}
\newcommand{\deltaTLone}{\ensuremath{\delta_{T_{\Lone}}}}
\newcommand{\MdeltaM}{\ensuremath{\bf{\vM_{\delta_M}}}}
\newcommand{\deltaM}{\ensuremath{\bm{\delta}_M}}
\newcommand{\deltaMBL}{\ensuremath{\bm{\delta}_{M_{BL}}}}
\newcommand{\deltaMLone}{\ensuremath{\bm{\delta}_{M_{\Lone}}}}

\newcommand{\Ident}{\ensuremath{\mathbb{I}}}

\newcommand{\Lone}{\ensuremath{\mathcal{L}_1}}
\newcommand{\transpose}{\ensuremath{\mathsf{T}}}

\newcommand{\greencheck}{{\color{green}{\ding{51}}}}
\newcommand{\redx}{{\color{red}{\ding{55}}}}

\newtheorem{remark}{Remark}

\def\BibTeX{{\rm B\kern-.05em{\sc i\kern-.025em b}\kern-.08em
    T\kern-.1667em\lower.7ex\hbox{E}\kern-.125emX}}
\begin{document}

\title{\LARGE \bf  $\Lone$-Adaptive MPPI  Architecture for Robust and Agile Control of Multirotors}

\author{Jintasit Pravitra$^{1*}$, Kasey A. Ackerman$^{2*}$, Chengyu Cao$^{3}$, Naira Hovakimyan$^{2}$, and Evangelos A. Theodorou$^{1}$
\thanks{$^{1}$Jintasit Pravitra and Evangelos A. Theodorou are with the School of Aerospace Engineering, Georgia Institute of Technology, Atlanta, GA 30332, USA {\tt\small jpravitra, evangelos.theodorou@gatech.edu}}
\thanks{$^{2}$Kasey A. Ackerman and Naira Hovakimyan are with the Mechanical Science and Engineering
Department, University of Illinois at Urbana-Champaign, Urbana, IL 61801 {\tt\small kaacker2, nhovakim@illinois.edu}}
\thanks{$^{3}$Chengyu Cao is with the Department of Mechanical Engineering, University of Connecticut, Storrs, CT 06269 {\tt\small chengyu.cao@uconn.edu}}
\thanks{$^{*}$equal contribution}
        }
\maketitle
\begin{abstract}
This paper presents a multirotor control architecture, where Model Predictive Path Integral Control (MPPI) and $\Lone$ adaptive control are combined to achieve both fast model predictive trajectory planning and robust trajectory tracking. MPPI provides a framework to solve nonlinear MPC with complex cost functions in real-time. However, it often lacks robustness, especially when the simulated dynamics are different from the true dynamics. We show that the $\Lone$ adaptive controller robustifies the architecture, allowing the overall system to behave similar to the nominal system simulated with MPPI. The architecture is validated in a simulated multirotor racing environment.
%without linearization
%everything remains nonlinear
\end{abstract}

\section{Introduction}
As multirotor aircraft continue to be integrated into our daily lives, researchers are challenged by the demand to automate complex tasks such as urban air transport, package delivery, autonomous racing, indoor exploration, or landing on a moving platform. These tasks often involve agile maneuvers and require complex trajectory planning. Despite recent advances, numerous challenges related to the problems of online optimal trajectory planning and replanning remain to be addressed.

Model predictive path integral control (MPPI) \cite{williams2016aggressive,williams2017information,williams2018information} offers a framework to efficiently solve a finite horizon nonlinear optimal control problem without restrictions on the form of the state cost function. The method is sampling based, and it leverages recent advances in GPU programming. Thousands of trajectories are propagated forward in parallel, and the optimal control is obtained by weight-averaging the controls from these trajectories (with each weight corresponding to the respective trajectory cost).

However, much like any simulation-based optimal control algorithm, MPPI suffers a degradation of robustness when the simulated dynamics are different from the true dynamics of the vehicle. In this case the true trajectory may diverge from the planned trajectory since the control sequence is optimized for the nominal dynamics but applied to the off-nominal plant dynamics.
The new initial state (for the purpose of MPC) can be significantly different from the previously planned state, thus rendering the warm-started controls to be far from optimal or in the worst case detrimental. 
A typical strategy to combat this issue is to include a tracking controller on top of the trajectory generator. For example in \cite{tubeMPPI}, iterative linear quadratic Gaussian (iLQG) is applied on top of MPPI to steer the states back to the planned trajectory. iLQG, however, requires a smooth cost function and also requires successive linearization of dynamics and cost function. Here we take a different approach and improve robustness through augmentation by an adaptive control element. 

\begin{figure}
    \centering\includegraphics[scale=0.45]{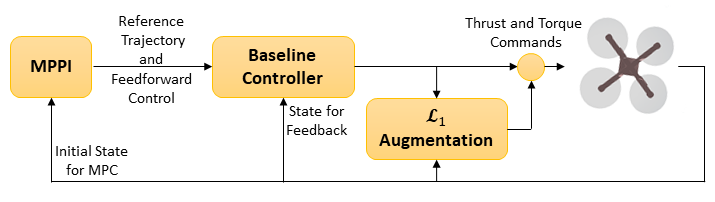} \caption{Overall Architecture}\label{fig:simplified_arch}
\end{figure}
Among the various adaptive control methods, $\Lone$ adaptive control \cite{L1book,hovakimyan20111} has been widely adopted due to its attractive properties of fast adaption, guaranteed robustness, and predictable transient response. These properties have been verified --- consistently with the theory --- in multiple manned and unmanned flight tests and simulations, on a diverse array of aircraft,
%\cite{ackerman2016l1,cotting2016can,puig2019l1,ackerman2019recovery,xargay2012l1,leman2009l1,choe2011handling,bichlmeier2016certifiable,peter2012l1}
% $\Lone$ adaptive control laws have been applied to a diverse array of aircraft, 
including fixed-wing aircraft~\cite{ackerman2016l1,cotting2016can,puig2019l1,ackerman2019recovery,xargay2012l1,leman2009l1,choe2011handling}, multirotors \cite{mallikarjunan2012l1,mallikarjunan2013L1,wang2013non}, helicopters \cite{bichlmeier2016certifiable}, and air-defense missiles \cite{peter2012l1}. 
% Another attractive property $\Lone$ adaptive control is its flexibility. 
The $\Lone$ adaptive control law can be formulated either as a standalone controller or as an augmentation of a baseline controller. Since adaptive control is a purely reactive control methodology, there is an opportunity to incorporate the advantages of baseline control algorithms that are capable of planning, such as MPC algorithms. 
\begin{figure*}
    \centering
    \vspace{2mm}
    \includegraphics[scale=0.45]{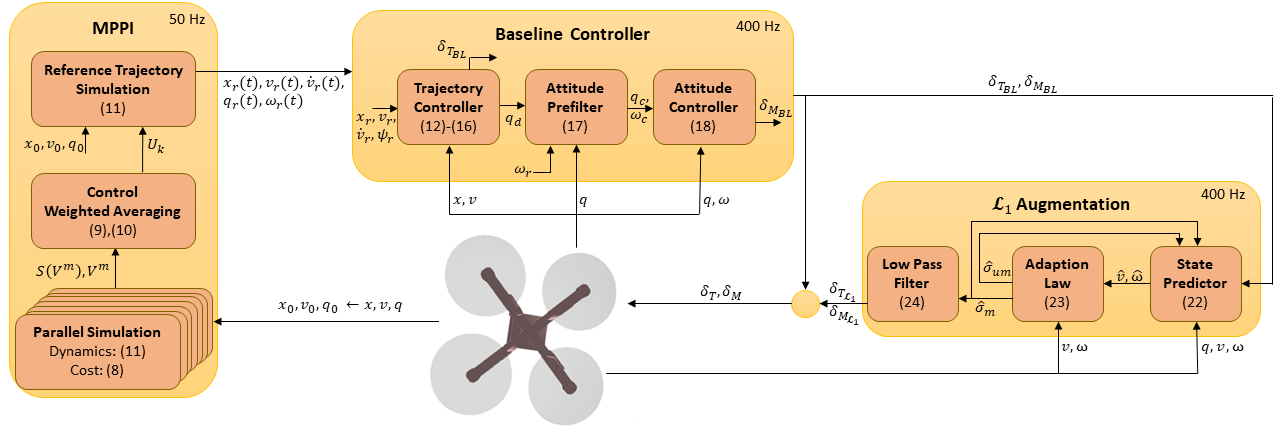}
    \caption{Detailed Architecture}
    \label{fig:detailed_arch}
    \vspace{-2mm}
\end{figure*}
Previously, in~\cite{pereida2018adaptive} and~\cite{pereidaall}, an $\Lone$ controller was used as an augmentation to a linear MPC, with a SISO linear reference model describing the desired behavior of the vehicle position response. While simpler to implement, this type of reference model does not take advantage of additional information available from the vehicle dynamics. By explicitly accounting for the vehicle dynamics, uncertainties can be directly compensated. To the best of the authors' knowledge however, there has never been an architecture where nonlinear MPC with complex cost function is robustified using $\Lone$ adaptive augmentation, which is the contribution of this paper. 
% We make use of the richness of structure in the $\Lone$ control law to define a nonlinear reference model defined by a geometric trajectory-tracking control law. The adaptive controller acts as an augmentation to the geometric control law, while an MPPI control law provides 
%This paper is organized as follows....
\section{Architecture}
The architecture --- depicted in Fig. \ref{fig:simplified_arch} --- consists of three components. The first component is MPPI, which acts as a nonlinear MPC controller generating the optimized trajectory and the feedforward control sequence. Due to the recent advances in GPU programming, MPPI can replan at a rate as fast as 50 Hz. The second component is the baseline controller, which provides some robustification by recomputing the attitude command to correct for position and velocity deviation. The attitude command modificiation is obtained through geometric control for multirotor dynamics with attitude tracking achieved through a quaternion-based PD controller. The third component is the $\Lone$ adaptive augmentation. By defining the state predictor to be the same as the MPPI nominal dynamics, the $\Lone$ augmentation  compensates for the mismatch between the nominal dynamics and the true dynamics. Hence, the tuning effort on the actual hardware can be reduced. A detailed  architecture with all sub-components and signal flows can be found in Fig. \ref{fig:detailed_arch}. 

\section{Multirotor Dynamics}
In this section, we describe the multirotor dynamics and quantify its (lumped) uncertainties.
Let the inertial basis vectors $\ve_1,\ve_2,\ve_3$ be aligned with the North-West-Up directions. Define the rotation matrix from the body frame to the inertial frame to be $\vR_B^I \triangleq \begin{bmatrix} \vb_1 & \vb_2 & \vb_3 \end{bmatrix}$. Define the attitude kinematics matrix to be 
\begin{align}
    \bm{\Omega}(\vomega) \triangleq \frac{1}{2}\begin{bmatrix} 0 &\ -\omega_1 &\ -\omega_2 &\ -\omega_3 \\
    \omega_1 &\ 0 &\ \omega_3 &\ -\omega_2 \\ 
    \omega_2 &\ -\omega_3 &\ 0 &\ \omega_1\\
    \omega_3 &\ \omega_2 &\ -\omega_1 &\ 0
    \end{bmatrix}.
\end{align}
The multirotor dynamics, with state vector $\vX = \begin{Bmatrix} \vx, & \vv, & \vq, & \vomega \end{Bmatrix}$ and control vector $\delta = \begin{Bmatrix} \deltaT & \deltaM \end{Bmatrix}$, are 
\begin{align}
\dot{\vx} &= \vv \qquad &
\bar{m}\dot{\vv} &=  \vb_3 \bar{T}_{\delta_T} \deltaT -\bar{m}g\ve_3 + \bar{\vzeta}\\
\dot{\vq} &= \bm{\Omega}(\vomega) \vq  \qquad &
\bar{\vJ}\dot{\vomega} &= \bar{\vM}_{\delta_M} \deltaM - \vomega\times\bar{\vJ}\vomega + \bar{\vxi} \nonumber
\end{align}
%where $\bar{\vzeta}$ and $\bar{\vxi}$ are lumped uncertainty forces and moments encapsulate exogenous disturbance forces, model uncertainty, and unmodeled dynamics.
where $\bar{\vzeta}$ and $\bar{\vxi}$ are unknown forces and moments which encapsulate exogenous disturbance forces, model uncertainty, and unmodeled dynamics.
Such uncertainties may include --- but are not limited to --- unknown aerodynamics, misalignment of the mass center, and parametric uncertainty. The parameters $\bar{m}, \bar{\vJ},\bar{T}_{\deltaT}, \bar{\bm{M}}_{\deltaM}$, corresponding to mass, moment of inertia, thrust control power, and moment control power respectively, are unknown in practice and only their nominal values
$m$, $\vJ$, $\TdeltaT$, $\MdeltaM$ are available. The multirotor dynamics can be rewritten with the nominal parameters and the lumped uncertainties ($\vzeta$ and $\vxi$) as 
\begin{align}
\dot{\vx} &= \vv \label{eq.xdot} \\
\dot{\vv} &=  \frac{\TdeltaT \deltaT}{m}\vb_3 -g\ve_3 + \frac{\TdeltaT}{m} \vR_B^I \vzeta \label{eq.vdot} \\
\dot{\vq} &= \bm{\Omega}(\vomega) \vq \label{eq.qdot} \\
\dot{\vomega}  &= \vJ^{-1}\MdeltaM \deltaM + \vJ^{-1}\MdeltaM\vxi \label{eq.omegadot}
\end{align}
with $\vzeta = \frac{m}{\bar{m} \TdeltaT}\vR_I^B\bar{\vzeta} + \left(\frac{m\bar{T}_{\deltaT}}{\bar{m}\TdeltaT}-1\right)\vR_I^B\vb_3\deltaT$ and $\vxi = \vM_{\delta_M}^{-1}\vJ\bar{\vJ}^{-1}\bar{\vxi} \ +  \left(\vM_{\deltaM}^{-1}\vJ\bar{\vJ}^{-1}\bar{\vM}_{\deltaM} -\Ident\right)\deltaM  + \vM_{\deltaM}^{-1}\vJ\bar{\vJ}^{-1}\left( \vomega \times \bar{\vJ}\vomega\right).$
A similar uncertainty parameterization can be found in \cite{mallikarjunan2012l1}.

\section{MPPI Trajectory Generation}

\subsection{Review of MPPI}
In this section, we provide a brief review of MPPI. Readers are encouraged to see \cite{williams2017model} for the stochastic Hamilton-Jacobi-Bellman derivation, \cite{williams2016aggressive,williams2017information,williams2018information} for the information theoretic derivation, and \cite{wang2019information,boutselis2019constrained} for the stochastic optimization derivation. In this paper, we use the information theoretic derivation.

Consider the discrete-time dynamics $\vs_{t+1} = \vF(\vs_t, \vv_t) \label{eq:mppi_dyn}$, where $\vv_t$ is a Gaussian distributed random control input $\vv_t \sim \mathcal{N}(\vu_t,\Sigma)$. We are interested in finding the mean control sequence $\vU = \begin{Bmatrix} \vu_0 , \vu_1, \dots \vu_{T-1}\end{Bmatrix}$ that minimizes the cost function 
\begin{align}
     J(\vU) =  \mathbb{E} \Bigg[ \ \phi(\vs_T) +  \sum_{t=0}^{T-1} \Big[ Q(\vs_t) + \lambda \vu^\transpose_t \Sigma^{-1}  \vu_t \Big] \Bigg].
\end{align}
%is minimized. 
Note that the state running cost, $Q(\vs_t)$, and the terminal cost,~$\phi(\vs_T)$, can be arbitrarily complex. For each control sequence realization $\vV^m = \begin{Bmatrix} \vv^m_0 , \vv^m_1, \dots \vv^m_{T-1} \end{Bmatrix}$, define $S(\vV^m)$ to be the state-dependent portion of the cost
\begin{align}
S(\vV^m) = \phi(\vs^m_T) + \sum_{t=0}^{T-1} Q(\vs^m_t).
\end{align}
Start from the initial condition $\vs_0$. Given the control sequence from the previous iteration $\vU_{k-1}$, simulate (thousands of) trajectories in parallel, each with a different control sequence realization $\vV^m$. The costs are collected for each rollout and are mapped to the trajectory weights:
\begin{align}
w(\vV^m) = \exp \Big( -\frac{1}{\lambda} \Big( S(\vV^m) - \sum_{t=0}^{T-1} \vu_{t,k-1}^\transpose \Sigma^{-1} \vv^m_t - \rho \Big) \Big) \label{eq.trajweight}
\end{align}
 The term $\rho$ is included in order to prevent arithmetic underflow. The value of $\rho$ is typically set to the minimum cost among all sampled trajectories. This term does not affect the solution because of the normalization (as will be seen in the next step). The optimal control at each time step can be approximated as
\begin{align}
    \vu_{t,k} = \vu_{t,k-1} +\frac{1}{\sum_{m=1}^M w(\vV^m)} \sum_{m=1}^M\Big[w(\vV^m) \epsilon^m \Big]
    \label{eq.optcontrol}
\end{align}
where $\epsilon^m \sim \mathcal{N}(0,\Sigma)$ and $M$ is the number of samples.

\subsection{Multirotor Application}
To apply MPPI to a multirotor trajectory generation problem, we propagate the kinematic equations and the disturbance-free translational dynamics. The rotational dynamics are neglected, as it is more appropriate to handle the rotational dynamics at the low-level controller. The sampled control inputs are thrust and angular velocity: $\vv_t = [\delta_T \ u_{\omega_1} \ u_{\omega_2} \ u_{\omega_3} ]^\transpose$.

However, injecting discontinuous angular velocity is not a realistic representation of the low-level controller. To represent this effect, the sampled angular velocity is low-pass filtered before it enters the attitude kinematics. In summary, we define the MPPI states to be $s = \begin{Bmatrix} \vx, \vv, \vq, \omega \end{Bmatrix} $ and propagate the trajectories using the following equations:
\begin{align}
\dot{\vx} &= \vv  \label{eq.xdotmppi} \qquad &
\dot{\vv} &=  \frac{\TdeltaT \deltaT}{m}\vb_3 -g\ve_3  \\
\dot{\vq} &= \bm{\Omega} \Big([\omega_1 \ \omega_2 \ \omega_3]^\transpose \Big) \vq \qquad &
\dot{\omega}_i &= \frac{u_{\omega_i} - \omega_i}{\tau_{\omega_i}} , \ i=1,2,3 \nonumber
\end{align}
After the trajectories are propagated and the optimal control sequence is obtained through (\ref{eq.trajweight}) and (\ref{eq.optcontrol}), the optimal control sequence is simulated through (\ref{eq.xdotmppi}) to obtain the reference trajectory $\begin{Bmatrix} \vx_r(t), & \vv_r(t), & \vq_r(t), & \vomega_r(t) \end{Bmatrix}$. Additionally, the reference acceleration sequence $\dot{\vv}_r(t)$ is also saved to be used in the baseline controller. 

\section{Baseline Controller}
We provide the procedure to obtain the baseline control input $\deltaTBL$ and $\deltaMBL$. The procedure follows the geometric control methods as in \cite{mellinger2011minimum,faessler2015automatic}, and \cite{lee2010geometric}.
\subsection{Baseline Trajectory Tracking Controller}
The desired specific force includes the feedforward term $\dot{\vv}_r$ from MPPI and feedback terms to correct for position and velocity deviation:
\begin{align}
    \vf_d = \sat_{\va_{max}}[ \dot{\vv}_r + \vK_P (\vx_r - \vx) + \vK_D (\vv_r - \vv)] + g\ve_3 \label{eq:linaccel_fb},
\end{align}
where $\sat[\cdot]$ is a function that clamps the acceleration to be between $[-\va_{max},\va_{max}]$. The baseline throttle command is proportional to the norm of the specific force:
\begin{align}
    \deltaTBL = (m/\TdeltaT) ||\vf_d||
\end{align}
Define the desired rotation matrix to be $\vR_D^I \triangleq \begin{bmatrix} \vd_{1} & \vd_{2} & \vd_{3} \end{bmatrix}$. For a vehicle where all rotors are aligned in a single plane, $\vf_d$ must be aligned with $\vd_{3}$:
\begin{align}
    \vd_{3} = \vf_d/||\vf_d||
\end{align}
We obtain the reference heading $\psi_r$ from $\vq_r$ and constrain the local level projection of the vehicle's nose to be \begin{align}
    \vl_{1} = [\cos \psi_r \ \ \sin \psi_r \ \ 0]^\transpose.
\end{align}
The other two components of the $\vR_D^I$
can be obtained using the following relationships:
\begin{align}
\vd_{2} = (\vd_{3} \times \vl_{1})/||\vd_{3} \times \vl_{1}||  \qquad
\vd_{1} =  \vd_{2} \times \vd_{3} \label{eq:desired_rotmat}
\end{align}
At this point, the desired attitude $\vR_D^I$ is completely defined.
\subsection{Baseline Attitude Controller}
Define $\vq_d$ to be the quaternion representation of $\vR_D^I$, and define $\vR_R^I$ to be the rotation matrix representation of $\vq_r$. Since MPPI and the baseline controller run at different rates, the calculation (\ref{eq:linaccel_fb}) - (\ref{eq:desired_rotmat}) may result in discontinuous attitude command. The following prefilter is recommended in order to smooth the command
\begin{align}
    \dot{\vq}_c     &= \bm{\Omega}(\vomega_c) \dot{\vq}_c \\
    \dot{\vomega}_c &= \vK_\omega \big[ \sat_{\vomega_{max}}[\vK_\omega^{-1} \vK_q \widetilde{\bf{Q}}(\vq_d,\vq_c)] +  ( \vR_I^B \vR_R^I \vomega_r - \vomega_c) \big], \nonumber
\end{align}
where $\vq_c$ and $\vomega_c$ are the filtered quaternion and angular velocity, and $\widetilde{\bf{Q}}(\cdot,\cdot)$ is the quaternion error function \cite{johnson2000limited}. Once the filtered state is obtained, it can be used in the baseline PD attitude controller:
\begin{align}
    \deltaMBL = \vK_\omega \big[ \sat_{\vomega_{max}}[\vK_\omega^{-1} \vK_q \widetilde{\bf{Q}}(\vq_c,\vq)] +  (\vomega_c - \vomega) \big]
\end{align}
The prefilter and the attitude controller gains ($\vK_q$ and $\vK_\omega$) are selected to be the same in this case, but this is not a requirement of the algorithm.

\section{$\Lone$ Adaptive Augmentation}
The $\Lone$ augmentation implements a nonlinear reference model~\cite{wang2012l1}, with estimations of both matched and unmatched uncertainties~\cite{xargay20101}.
%follow the formulation found in~\cite{xargay20101}, implementing a nonlinear reference model as in~\cite{wang2012l1}. 
%The adaption law estimates both matched and unmatched uncertainties. 
The unmatched uncertainty arises in the translational dynamics, since a multirotor (with rotors aligned in a single plane) can only provide linear acceleration along the body z-axis. Because the uncertainties~$\vzeta$,~$\vxi$, appear purely in the vehicle dynamics, the kinematic equations can be omitted and consideration given only to the dynamic equations~\eqref{eq.vdot} and~\eqref{eq.omegadot}, which are rewritten as 
\begin{align}
\dot{\vv} &= -g\ve_3 + \frac{\TdeltaT}{m}\vb_3 (\deltaTBL + \deltaTLone + \zeta_m) \notag \\
&\hspace{10mm}  + \frac{\TdeltaT}{m}\begin{bmatrix} \vb_1 & \vb_2 \end{bmatrix}\vzeta_{um} \\
\dot{\vomega}  &= \vJ^{-1}\MdeltaM( \deltaMBL + \deltaMLone + \vxi_m) ,
\end{align}
where substitutions have been made for 
\begin{align*}
\vzeta = \begin{bmatrix} \vzeta_{um} \\ \zeta_m\end{bmatrix}, \quad
\deltaT = \deltaTBL+\deltaTLone, \quad
\deltaM = \deltaMBL+\deltaMLone.
\end{align*}
In a more general form, the dynamics can be written as
\begin{equation}
 \dot{\bm{z}} = \vf\left(\vR_B^I\right) + \vg\left(\vR_B^I\right)\left(\vu_{\Lone} + \vsigma_m\right) + \vg^\perp\left(\vR_B^I\right)\vsigma_{um}
\end{equation}
where 
\begin{align*}
\bm{z} &= \begin{bmatrix}\vv^\transpose & \vomega^\transpose \end{bmatrix}^\transpose \qquad
& \vu_{\Lone} &= \begin{bmatrix} \delta_{T_{\Lone}} & \delta_{\vM_{\Lone}}^\transpose \end{bmatrix}^\transpose \\
\vsigma_m &= \begin{bmatrix} \zeta_m & \vxi^\transpose \end{bmatrix}^\transpose
& \vsigma_{um} &= \vzeta_{um} &
\end{align*}
\begin{align*}
\vf(\vR_B^I) &= \begin{bmatrix} -g\ve_3 + m^{-1}\vb_3 \TdeltaT\deltaTBL \\ \vJ^{-1}\MdeltaM\deltaMBL \end{bmatrix}
\\
\vg(\vR_B^I) &= \begin{bmatrix} m^{-1}\vb_3\TdeltaT & \bm{0}_{3\times 3} \\ \bm{0}_{3\times 1} & \vJ^{-1}\MdeltaM \end{bmatrix} \\
\vg^\perp(\vR_B^I) &= \begin{bmatrix}m^{-1}\vb_1\TdeltaT & m^{-1}\vb_2\TdeltaT \\ \bm{0}_{3\times 1} & \bm{0}_{3\times 1} \end{bmatrix}
\end{align*}
Then the $\Lone$ state predictor is defined as
\begin{equation}
 \dot{\hat{\bm{z}}}= \vf + \vg\left(\vu_{\Lone} + \hat{\vsigma}_m\right) + \vg^\perp\hat{\vsigma}_{um} + \vA_s\tilde{\bm{z}},
\end{equation}
where $\tilde{\bm{z}} = \hat{\bm{z}}-\bm{z}$ and $\vA_s$ is a Hurwitz matrix. For simplicity, we assume $\hat{\bm{z}}(0) = \bm{z}(0)$ and note that a non-zero state predictor initialization error results in an exponentially decaying term~\cite{cao2007effect}. Define $\bm{\Phi} \triangleq \vA_s^{-1}\left(\exp(\vA_sT_s) - \Ident \right)$. The piecewise-constant adaption law is
\begin{equation}
\begin{bmatrix} \hat{\vsigma}_m(iT_s) \\ \hat{\vsigma}_{um}(iT_s) \end{bmatrix} = -\begin{bmatrix} \Ident_{4\times 4} & \bm{0}_{4\times 2} \\ \bm{0}_{2\times 4} & \Ident_{2\times 2} \end{bmatrix} \vG(iT_s)^{-1}\bm{\Phi}^{-1} \bm{\mu}(iT_s)
\end{equation}where $T_s$ is the time step and where $
    \vG(iT_s) = \begin{bmatrix} \vg\left(\vR_B^I\right) \ \vg^\perp\left(\vR_B^I\right) \end{bmatrix}
$ and $
\bm{\mu}(iT_s) = \exp(\vA_sT_s) \tilde{\bm{z}}(iT_s) 
$ are evaluated at the i-th time step.
The $\Lone$ control law compensates only for the matched components of the uncertainty within the bandwidth of the strictly proper stable filter $\vC(s)$:
\begin{equation}
    \vu_{\Lone} = -\vC(s) \hat{\vsigma}_m.
\end{equation}
The effects of the unmatched component are estimated to improve the performance of the state predictor but, for this application, it is not necessary to compensate for the unmatched components directly; they can be indirectly canceled by the baseline control law. 
Proofs regarding stability and bounds on states and controls can be found in~\cite{wang2012l1}.

\begin{remark}
	Define the (invertible) elementary column operator 
	\begin{equation}
	\vE = \begin{bmatrix}
	0 & 0 & 1 & 0 & 0 & 0 \\
	0 & 0 & 0 & 1 & 0 & 0 \\
	0 & 0 & 0 & 0 & 1 & 0 \\
	0 & 0 & 0 & 0 & 0 & 1 \\
	1 & 0 & 0 & 0 & 0 & 0 \\
	0 & 1 & 0 & 0 & 0 & 0
	\end{bmatrix} 	
	\end{equation}
	and let $\vH = \vG \vE $. Then 
	\begin{equation}
	\vH = \begin{bmatrix} m^{-1}\vR_B^I\TdeltaT & \bm{0}_{3\times 3} \\ \bm{0}_{3\times 3} & \vJ^{-1}\MdeltaM\end{bmatrix}
	\end{equation}
	The inverse of $\vH$ is trivially calculated as
	\begin{equation}
	\vH^{-1} = \begin{bmatrix} m\vR_I^B\TdeltaT^{-1} & \bm{0}_{3\times 3} \\ \bm{0}_{3\times 3} & \vJ\vM_{\deltaM}^{-1} \end{bmatrix},
	\end{equation}
	from which we can directly obtain 
	\begin{equation}
	\vG^{-1} = \vE\vH^{-1} = \begin{bmatrix}
	m\vb_1^\transpose\TdeltaT^{-1} & \bm{0}_{1\times 3} \\
	\bm{0}_{3\times 3} & \vJ\vM_{\deltaM}^{-1} \\
	m\vb_2^\transpose\TdeltaT^{-1} & \bm{0}_{1\times 3} \\
	m\vb_3^\transpose\TdeltaT^{-1} & \bm{0}_{1\times 3}
	\end{bmatrix}
	\end{equation}
	Since $\TdeltaT$ and $\MdeltaM$ are known and the rotation matrix $\vR_B^I$ is obtained from sensor/estimator feedback, the nullspace component $\vg^\perp$ can be directly obtained as well.
	Therefore, even though the span of $\vg$ and the span of its nullspace are changing in time, no matrix inversion operations are needed to compute either $\vg^\perp$ or $\vG^{-1}$.
 \end{remark}

Since the $\Lone$ adaptive augmentation is intended to be implemented in a discrete-time environment, we formulate the discrete version of the adaptation law as
\begin{equation}
\begin{bmatrix} \hat{\vsigma}_{m,k} \\ \hat{\vsigma}_{um,k} \end{bmatrix} = -\begin{bmatrix} \Ident_{4\times 4} & \bm{0}_{4\times 2} \\ \bm{0}_{2\times 4} & \Ident_{2\times 2} \end{bmatrix} \vG_k^{-1}\bm{\Phi}^{-1}\bm{\mu}_k 
\end{equation}
Here, $(\cdot)_{k}$ is equivalent to $(\cdot)(kT_s)$. The discrete version of the control law can be written as
\begin{align}
\vx_{f,k+1} = \vA_f \vx_{f,k} + \vB_f\hat{\vsigma}_{m,k} \qquad
\vu_{\Lone,k} = -\vC_f\vx_{f,k} \nonumber
\end{align}
where $\{\vA_f,\vB_f,\vC_f\}$ are the discrete-time state space matrices defining the low-pass filter $\vC(s)$ and $\vx_{f}$ is the filter state.
The discrete-time state predictor is propagated via
\begin{align}
 \hat{\bm{z}}_{k+1} &= \hat{\bm{z}}_k + \left[\vf_k + \vg_k\left(\vu_{\Lone,k} + \hat{\vsigma}_{m,k}\right)\right. \notag \\
  &\qquad  + \left. \vg_k^\perp\hat{\vsigma}_{um,k} + \vA_s\tilde{\bm{z}}_k \right] T_s,
\end{align}
where $\tilde{\bm{z}}_k = \hat{\bm{z}}_k - \bm{z}_k$.

\section{Results}
\begin{figure}[h!]
    \centering
    \includegraphics[width=0.7\textwidth]{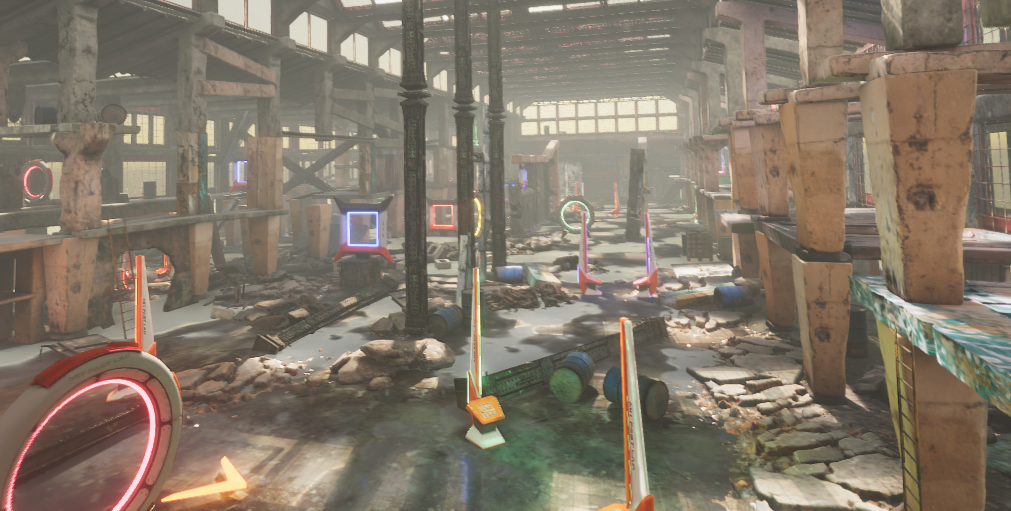}
    \caption{FlightGoggles Racing Environment}
    \label{fig:fgsreenshot}
    \vspace{-0mm}
\end{figure}

The control architecture is evaluated in FlightGoggles, a photorealistic quadrotor racing simulation environment based on Unity and ROS \cite{FlightGoggles}. The objective is to fly through the designated gates in the correct sequence and finish the course as fast as possible. We assume that the gate locations are known a priori and that the state estimation is perfect. A screenshot of FlightGoggles is shown in Fig. \ref{fig:fgsreenshot}.  
\subsection{Cost Function and Parameters}
\begin{figure}[h!]
        \centering
        \includegraphics[width=0.8\textwidth]{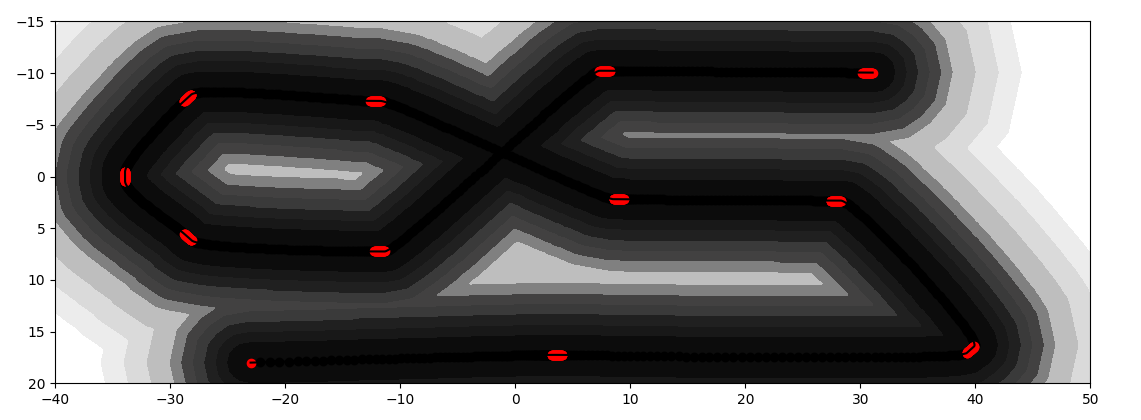}
        \caption{Top view of the air corridor cost map}
        \label{fig:costmap}
        %\vspace{-2mm}
    \end{figure}

The MPPI cost function is designed to aggregate the following objectives:
\begin{itemize}
    \item maintain the (3-dimensional) air corridor's centerline;
    \item point the front of the multirotor toward the next gate;
    \item maintain the commanded speed;
    \item heavily penalize trajectories that violate the air corridor;
    \item give a bonus to trajectories that passes through gates.
\end{itemize}
The expression of the cost function is 
\begin{multline}
Q(\vs) = 450 M(\vx) + 250 |\text{mod}(\psi_{cmd} - \psi)| \\ + 150 \Big| V_{cmd} - \sqrt{v^2_x + v^2_y} \Big|  + 10000\mathds{1}_{out}(\vx)  - 150\sum \mathds{1}_{gate}(\vx),
\end{multline}
where $M(\vx)$ is the air corridor cost map look-up as shown in Fig. \ref{fig:costmap}. We select the planning horizon to be 1.5 s. Since the units of $\deltaT$ and $\deltaM$ are arbitrary, we can conveniently select $\TdeltaT = 1.0$ and $\MdeltaM = \Ident_{3\times 3}$ so that $\deltaT$ is in the unit of Newtons and $\deltaM$ is in the unit of Newton-meters. This can be done without loss of generality. The derivation of the lumped uncertainties is still valid with the scaled values of $\bar{T}_{\deltaT}$ and $\bar{\vM}_{\delta_M}$. The rest of the parameters are shown in Table~\ref{tab:params}.
\begin{figure}[t!]
\vspace{3mm}
    \centering
    \includegraphics[scale=0.25]{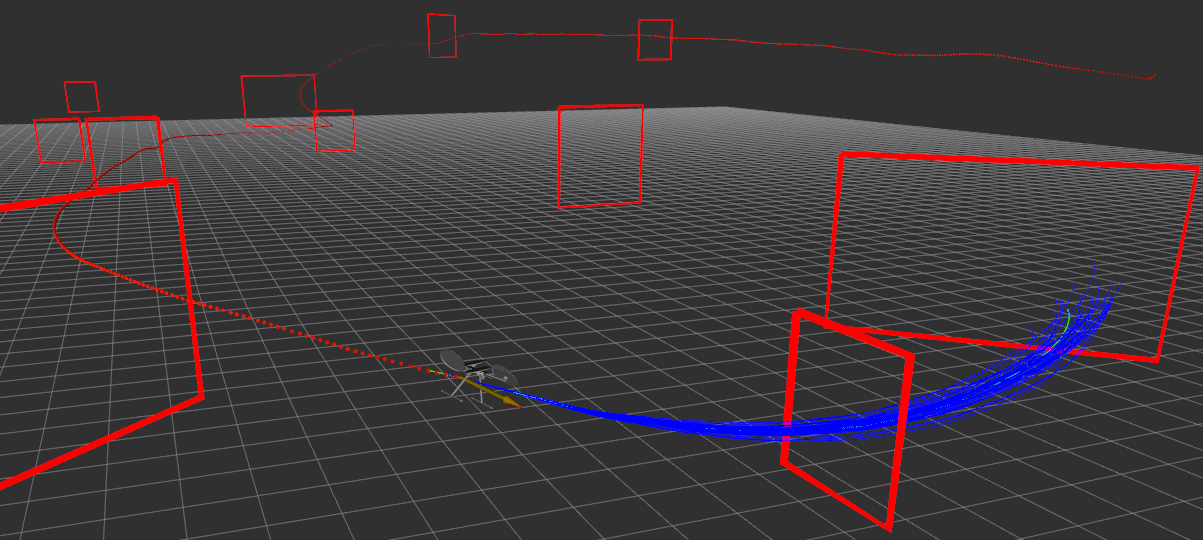}
    \caption{Ground station visualization\vspace*{6mm}}
    \label{fig:rviz_viz}
    \vspace{0mm}
\end{figure}
\begin{table}[h!]
\caption{Parameters}
\vspace{-3mm}
\begin{tabular}{|c|cc|}
\hline
\multicolumn{3}{|c|}{Nominal Plant} \\
\hline\hline
$m$ & $1.0$ & $kg$ \\
\hline
$\vJ$ & $\text{diag}(4.9,4.9,4.9) \times 10^{-3}$ & $kg \ m^2$ \\
\hline\hline
\multicolumn{3}{|c|}{MPPI} \\
\hline\hline
$\Sigma$ & $\text{diag}(\ 1.5\ , 0.4\ , \ 0.4 \ , \ 0.4 \ )$ & $N, rps$\\
\hline
M & 7,200 &\\
\hline
$\lambda$ & 1.4 &\\
\hline
$\tau_{\omega_1},\ \tau_{\omega_2},\ \tau_{\omega_3}$ & 0.25 & s\\
\hline
\hline
\multicolumn{3}{|c|}{Baseline Controller} \\
\hline\hline
$\vK_P$ & $\text{diag}(6.0,6.0,6.0)$  & \\
\hline
$\vK_D$ & $\text{diag}(4.0,4.0,4.0)$ & \\
\hline
$\vK_q$ & $\text{diag}(1.0,1.0,1.0)$ & \\
\hline
$\vK_\omega$ & $\text{diag}(0.15,0.15,0.15)$ &\\
\hline
$\va_{max}$ & $15.0$ & $m/s^2$\\ 
\hline
$\vomega_{max}$ & $2.0$ & $rps$\\ 
\hline
\hline
\multicolumn{3}{|c|}{$\Lone$ Controller} \\
\hline
\hline
$T_s$ &  2.5 & ms\\
\hline
$C(s)$ & $\frac{15}{(s+15)} \Ident_{4\times 4}$ & \\
\hline
$\vA_s$ & $-5.0 \Ident_{6\times 6}$ & \ \\
\hline
\end{tabular}
\label{tab:params}
\end{table} 

\subsection{Results}
\begin{figure}[h!]
\vspace{2mm}
    \centering
    \includegraphics[scale=0.6]{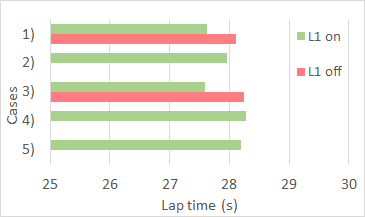}
    \caption{Average lap time}
    \label{fig:avlap}
    \vspace*{-2mm}
\end{figure}

We test our algorithm with and without $\Lone$ augmentation for the following cases:
\begin{enumerate}
    \item known dynamics model (since aerodynamic drag is not modeled in the nominal dynamics, some drag compensation is expected with $\Lone$ augmentation);
    \item mass increase by $50\%$;
    \item moment of inertia increase by $100\%$ in all axes;
    \item constant nose-up pitching moment disturbance of $0.1 \ N m$ (equivalent to center of gravity offset);
    \item reduction in motor thrust control power by $40\%$ (reduction in both  $\bar{T}_{\deltaT}$ and $\bar{\vM}_{\delta_M}$).
\end{enumerate}
Figure \ref{fig:rviz_viz} shows a visualization for one of the runs. For each case, we run the race 15 times. The average lap time is shown in Fig \ref{fig:avlap}. The bar graph is shown only if all 15 runs are successfully completed. (The vehicle neither crashes nor diverges.)  The fact that not all cases are successful is also shown in Table \ref{tab:success}.
\begin{table}[h!] 
 \caption{Success and failure for each case }
 \label{tab:success}
 \vspace{-3mm}
 \begin{tabular}{|c|c|c|} 
 \hline
 Case &  $\Lone$ off & $\Lone$ on \\ 
 \hline\hline
 1) &  \greencheck &  \greencheck \\ 
 2) & \redx & \greencheck \\
 3) & \greencheck & \greencheck \\
 4) & \redx & \greencheck \\
 5) & \redx & \greencheck\\
 \hline
\end{tabular}
\end{table}
Figure \ref{fig:avlap} and Table \ref{tab:success} show that $\Lone$ augmentation successfully reduces lap time. For the cases of extreme disturbances, $\Lone$ augmentation successfully ``saves" the vehicle that would have crashed if only the baseline controller was employed.
\vspace{3mm}

\section{Conclusion}
In this paper, we proposed a multirotor control architecture where MPPI acts as a nonlinear MPC controller while model uncertainty is compensated through an $\Lone$ adaptive controller. It was shown in the quadrotor racing scenario that inclusion of $\Lone$ augmentation improves performance and in many cases is imperative to the success of the mission. Potential future work includes flight testing, detailed analysis of robustness, and incorporation of the control limits. 

\section*{Acknowledgment}
The authors would like to thank the members of the Georgia Tech AlphaPilot team who initiated the implementation of MPPI with FlightGoggles, including Grady Williams, Sam Yi Ting, Jason Gibson, Manan Gandhi, Keuntaek Lee, Camilo Duarte, Akash Patel, and Matthew Houghton. The authors would also like to thank Aditya Gahlawat from UIUC for the fruitful discussion.

This work is funded by NASA LaRC.

\bibliographystyle{IEEEtran}
\bibliography{IEEEabrv,ep} 

\begin{thebibliography}{10}
\bibitem{pereidaall}
------, ``All models are wrong: Robust adaptive model predictive control for
  safe and high accuracy trajectory tracking in the presence of model errors.''
\end{thebibliography}
\end{document}